\documentclass[12pt]{article}
\usepackage[utf8]{inputenc}
\usepackage{graphicx}
\usepackage[english]{babel}
\usepackage{amsmath}
\usepackage{amssymb}
\usepackage[margin=1in]{geometry}
\usepackage{hyperref}
\usepackage{float}
\usepackage{multirow}
\usepackage{latexsym}
\usepackage{caption}
\usepackage{babelbib}
\usepackage{latexsym}
\usepackage[T1]{fontenc}
\usepackage{tabularx}
\usepackage{dcolumn}
\usepackage{eurosym}
\usepackage{setspace}
\usepackage{color}
\usepackage[usenames,dvipsnames]{xcolor}
\usepackage{geometry}
\usepackage{multicol}
\usepackage{caption}
\usepackage{subfigure}
\usepackage{amsthm}
\usepackage{stmaryrd}
\usepackage{enumitem}
\usepackage{physics}
\usepackage{bbold}
\usepackage[bb=ams]{mathalfa}

\newcommand{\im}[2]{{#1}_{\scriptscriptstyle{#2}}}
\newcommand{\irm}[2]{{#1}_{\scriptscriptstyle{\textrm{#2}}}}

\newcommand{\Ls}{\text{G}}
\newcommand{\Lf}{\text{KG}}

\newcommand{\dl}{d}

\newcommand{\sv}{b}
\newcommand{\svb}{(\sv)}
\newcommand{\sm}{\im{p}{b}}

\newcommand{\scpb}{(\sv,\sm)}
\newcommand{\fv}{\phi}
\newcommand{\fm}{\pi}
\newcommand{\svz}{\im{\sv}{0}}

\newcommand{\Hs}[1]{\irm{\mathcal{H}}{#1}}

\newcommand{\Hi}{\mathcal{H}}

\newcommand{\Ps}[1]{\im{\Gamma}{#1}}

\newcommand{\Sib}{\Sigma}

\newcommand{\FS}{KG}

\newcommand{\ol}{\im{\omega}{k}^{\scriptscriptstyle{(b)}}}

\newcommand{\f}{f^{(\sv)}}

\newcommand{\qs}[1]{\hat{#1}}
\newcommand{\qf}[1]{\pmb{#1}}

\newcommand{\h}[2]{\im{\qf{h}}{#1}^{#2}}

\newcommand{\heffscp}[1]{\im{\qf{h}}{\text{eff}#1}^{\scpb}}
\newcommand{\heff}[2]{\im{\qf{h}}{\text{eff}#1}^{#2}}

\newcommand{\ab}[2]{\im{\pmb{a}}{\vec{#1}{#2}}^{\scriptscriptstyle{(\sv)}}}
\newcommand{\abd}[2]{\left(\im{\pmb{a}}{\vec{#1}{#2}}^{\scriptscriptstyle{(\sv)}}\right)^{\!\dagger}}

\newcommand{\Us}[1]{\im{\qs{1}}{#1}}
\newcommand{\Uf}[1]{\im{\qf{1}}{#1}}

\newcommand{\ef}[1]{\im{e}{#1}^{\svb}}

\newcommand{\Epd}[1]{\im{E}{#1}^{\scpb}}

\newcommand{\en}[1]{\im{n}{\vec{#1},d}}
\newcommand{\efr}{\im{e}{\im{n}{d}}^{(\sv)}}

\newcommand{\EGsv}{\irm{E}{G}^{\scpb}}
\newcommand{\EG}{\irm{E}{G}}
\newcommand{\EKGsv}[1]{\im{E}{\mathrm{KG},#1}^{\scpb}}
\newcommand{\EKGr}{\im{E}{\mathrm{KG},n}}
\newcommand{\DE}[1]{\im{\Delta}{E,\vec{#1}}}
\newcommand{\Eone}{\im{E}{1,n,\vec{k}}^{\scpb}}
\newcommand{\Etwo}{\im{E}{2,n,\vec{k}}^{\scpb}}

\newcommand{\pib}[1]{\im{\qf{\pi}}{#1}^{(\sv)}}

\newcommand{\pipd}[1]{\im{\qf{\pi}}{#1}^{\scpb}}

\newcommand{\pli}[1]{\im{\pmb{\pi}}{#1}}
\newcommand{\prora}[1]{\efr \langle \im{e}{\im{n}{\vec{k},d}#1}^{(\sv)}, \cdot \rangle} 
\newcommand{\prola}[1]{\im{e}{\im{n}{\vec{k},d}#1}^{(\sv)} \langle \efr, \cdot \rangle} 
\newcommand{\prorf}[1]{\im{e}{\im{n}{d}}^{(\im{\sv}{0})} \langle \im{e}{\im{n}{\vec{k},d}#1}^{(\sv)}, \cdot \rangle} 
\newcommand{\prolf}[1]{\im{e}{\im{n}{\vec{k},d}#1}^{(\im{\sv}{0})} \langle \im{e}{\im{n}{d}}^{(\sv)}, \cdot \rangle} 
\newcommand{\proR}{\im{e}{\nd}^{(\im{\sv}{0})} \im{\langle \im{e}{\nd}^{(\im{\sv}{0})}, \cdot \rangle}{\Lf}} 

\newcommand{\us}[2]{\im{\pmb{u}}{#1}^{#2}}

\newcommand{\dr}{\mathrm{d}}

\newcommand{\sh}{\sqrt{^0\!\,h}}

\newcommand{\aone}[1]{\alpha_{1,\en{#1}}^{(\sv)}}
\newcommand{\atwo}[1]{\alpha_{2,\en{#1}}^{(\sv)}}

\newcommand{\pp}{\varepsilon}

\newcommand{\nd}{\im{n}{d}}
\newcommand{\nkd}{\im{n}{\vec{k},d}}
\newcommand{\rel}{\nd}
\newcommand{\refs}{\text{R}}

\let\OLDthebibliography\thebibliography
\renewcommand\thebibliography[1]{
  \OLDthebibliography{#1}
  \setlength{\parskip}{1.5pt}
  \setlength{\itemsep}{2pt plus 2ex}
}

\title{{\sf Quantum Cosmological Backreactions III:}\\
 {\sf Deparametrised Quantum Cosmological Perturbation Theory}} 
\author{
{\sf S. Schander}$^1$\thanks{{\sf 
susanne.schander@gravity.fau.de}},
{\sf T. Thiemann}$^1$\thanks{{\sf 
thomas.thiemann@gravity.fau.de}}\\
\\
{\sf $^1$ Inst. for Quantum Gravity, FAU Erlangen -- N\"urnberg,}\\
{\sf Staudtstr. 7, 91058 Erlangen, Germany}\\
}
\date{{\small\sf \today}}

\begin{document}

\maketitle

{\sf

\begin{abstract}
This is the third paper in a series of four in which we use space adiabatic methods in order to incorporate backreactions among the homogeneous and between the homogeneous and inhomogeneous degrees of freedom in quantum cosmological perturbation theory.

In this paper we consider a particular kind of cosmological perturbation theory which starts from a gauge fixed version of General Relativity. The gauge fixing is performed using a material reference system called Gaussian dust. The resulting system has no constraints any more but possesses a physical Hamiltonian that drives the dynamics of both geometry and matter. As observable matter content we restrict to a scalar field
(inflaton). We then explore the sector of that theory which is purely homogeneous and isotropic with respect to the geometry degrees of freedom but contains inhomogeneous perturbations up to second order of the scalar field.

The purpose of this paper is to explore the quantum field theoretical challenges of the space adiabatic framework in a cosmological model of inflation which is technically still relatively simple. We compute the quantum backreaction effects from every energy band of the inhomogeneous matter modes on the evolution of the homogeneous geometry up to second order in the adiabatic parameter. These contributions turn out to be significant due to the infinite number of degrees of freedom and are very sensitive 
to the choice of Fock representation chosen for the inhomogeneous matter
modes.
\end{abstract}

\newpage

\tableofcontents

\newpage

\section{Introduction}
\label{s1}

In a previous paper of this series \cite{1} we have emphasised 
the importance of an adequate description of backreaction effects 
between homogeneous and inhomogeneous degrees of freedom and between 
geometry and matter in quantum cosmology
\cite{2}. We have argued that the framework of space adiabatic perturbation
theory (SAPT) \cite{3} is ideally suited to do this because 
it is able to combine the framework of quantum field theory on 
curved classical spacetimes (QFT in CST) \cite{5} with the fundamental 
quantum nature of that background. The way this works is quite similar 
to the quantum field theory on non commutative spacetimes approach of \cite{6}  
where Weyl quantisation techniques are used in order to employ 
the ordinary QFT framework while spacetime coordinates are non-commutative.
Here, instead of the spacetime coordinates we use Weyl quantisation techniques
in oder to treat a non-commutative quantum background. Here the homogenous
degrees of freedom play the role of the ``slow'' degrees of freedom while the 
inhomogenous ones are ``fast''.  

However, SAPT not only harmonically synthesises the apparently contradicting
natures of the background geometry in quantum cosmology on the one hand 
and QFT in CST on the other but also provides a concrete scheme for how 
to systematically compute backreaction effects between the homogeneous quantum 
background geometry and the inhomogeneous quantum matter and quantum geometry. 
The separate treatment of the homogeneous and inhomogeneous degrees 
of freedom of a quantum field is called the hybrid scheme \cite{7} which 
we adopt in this series of papers. Moreover, the SAPT scheme can also deal
with the situation that the coupling between the slow and fast degrees of 
freedom depends on both, configuration and momentum variables, of the 
slow sector which goes beyond what one can do in a Born-Oppenheimer 
approach \cite{8}. More generally, it is able to treat the situation 
that the slow sector couples with mutually non-commuting operators.
As emphasised in \cite{9}, this is very important when one goes beyond 
cosmological
perturbation theory and tries to quantise geometry with the methods 
of Loop Quantum Gravity (LQG) \cite{10} while matter is quantised using 
the QFT in CST framework because in LQG even the spatial geoemetry to which 
matter fields couple becomes non-commutative.   

In this paper we consider General Relativity coupled to an inflaton field
as well as Gaussian dust \cite{11}. We use the dust fields to deparametrise 
the theory \cite{12} so that all geometry degrees of freedom (scalar, 
vector, tensor) and the inflaton become (Dirac) observables and the 
theory is equipped with a true conservative Hamiltonian which for 
Gaussian dust is particularly simple: it is nothing
but the gravitational and inflaton contribution to the Hamiltonian 
constraint integrated over space. One can then apply classical cosmological 
perturbation theory to this system in the fashion outlined in all detail in 
\cite{13} for a different material reference system \cite{14}. For the 
illustrative purpose of this paper we discard the inhomogeneous geometry 
degrees of freedom and focus on the homogeneous geometry and homogeneous 
and inhomogeneous inflaton sector to second order in the
inhomogeneities. Note that the inhomogeneous field species decouple 
to second order. We reserve the treatment of inhomogeneous geometry 
degrees of freedom to our fourth paper \cite{15}. This artificial 
restriction serves the purpose of mathematical convenience: We would 
like to illustrate the SAPT formalism in a cosmological, quantum field
theoretical context which is as simple as possible in order not 
to get lost in details that have nothing to do with the SAPT programme 
itself. We stress that the inhomogeneous geometry modes can be trteated by
exactly the same methods. 

The application of the SAPT scheme to this quantum field theoretical model 
in principle proceeds just like the purley homogeneous quantum mechanical
model \cite{4} with two important differences: First, it is absolutely
crucial that one first applies a canonical transformation (exact up 
to second order in the inhomogenous perturbations) to this system
in order to switch to variables which have the property that in terms of them
the background dependent Fock representations of the inflaton all employ
the same Hilbert space for every background. This is otherwise 
not the case as observed in \cite{9} where the very same model was considered
and which would then prevent the application of the SAPT scheme. In \cite{9}
therefore an artificial  
cut-off on the number of degrees of freedom had to be introduced.
That this can be avoided by means of a transformation which 
mixes homogeneous and inhomogeneous degrees of freedom was first observed 
in \cite{16} in a different context. Second, it is necessary to perform a 
further canonical 
transformation in the homogeneous sector in order to avoid the problems
that come from {\it tachyonic} quantum fields \cite{1}. After these
subtleties have been dealt with, we proceed as in \cite{4} and 
compute the second order correction 
(with respect to the adiabatic parameter, to be 
distinguished from the order with respect to the inhomogeneous perturbations)
to the inhomogeneous dynamics from every energy 
band of the inhomogenous sector. This adiabatic correction is highly 
non-trivial and is quite sensitive to the Fock representation
chosen in the inhomogeneous sector: If not carefully selected, the adiabatic
correction can easily diverge. These findings have potentially observational 
consequences since they should influence the details of the quantum 
cosmological bounce obeserved for instance in Loop Quantum Cosmology
(LQC) \cite{17} which describes the truncation of LQG to the homogeneous 
sector {\it without} backreactions.\\  
\\
The architecture of this paper is as follows:\\
\\
In section two we briefly introduce the model and prepare it for the 
application of the SAPT scheme by performing the afore mentioned field 
truncations and canonical transformations.

In section tree we then directly apply the  
SAPT framework. We assume the reader to be familiar with the notation and 
main formulae of \cite{1}. The obtained adiabatic corrections display
a rather singular character with respect to the homogeneous degrees 
of freeedom, however, the corresponding operator has, in the Schr\"odinger
representation, the computationally 
convenient dense and invariant domain of \cite{18}. In section four we summarise and conclude.

\section{The Hamiltonian}
The Hamiltonian, after a transformation which is canonical up to second order in the cosmological perturbations (see \cite{1} for more details), includes an effective mass term $M^2$ which for arbitrary phase space variables $(a,\im{p}{a})$ needs not to be positive definite. As already commented in \cite{1}, a possible solution for this, is to perform a canonical transformation to new cosmological variables $(\sv,\sm)$, given by,
\begin{equation} \label{eq:Definition slow variables}
a = \sqrt{\sv^2 + \sigma^2 \frac{\sm^2}{\sv^2}} =: \Sigma^{\scriptscriptstyle{\scpb}}, ~~~ \im{p}{a} = a \cdot \frac{\sm}{\sv}. 
\end{equation}
In order to not confuse the different sets of variables, we defined the scale factor $a$ as a function of $\scpb$ as $\Sigma$. According to space adiabatic perturbation theory, the symbol Hamiltonian which serves for the analysis in the sequel is thus given by, 
\begin{align}
\pmb{h} &= L^3 \left( - \frac{1}{12} \Sigma \frac{\sm^2}{\sv^2} + \Lambda\Sigma^3 \right)\Uf{\Lf} +\, \frac{1}{2 \Sigma} \!\int_{\mathbb{T}^3}\! \dr^3x \! \left( \frac{\pmb{\pi}^2}{\sh} + \sh\, \pmb{\phi} \left( -\!\vartriangle + \mu^2 b^2 \right) \pmb{\phi} \right) \nonumber \\
&=: \EGsv\,\Uf{\Lf} + \, \frac{1}{2 \Sigma} \!\int_{\mathbb{T}^3}\! \dr^3x \! \left( \frac{\pmb{\pi}^2}{\sh} + \sh\, \pmb{\phi} \left( -\!\vartriangle + \mu^2 b^2 \right) \pmb{\phi} \right), \label{eq:Hamilton Symbol}
\end{align}
where the space manifold is assumed to be a compact three-torus $\mathbb{T}^3$ with volume $L^3$. The commutation relations for the new cosmological pair and for the scalar field are,
\begin{equation}
\im{\left[\qs{\sv},\im{\qs{p}}{\sv} \right]}{\Ls} = \frac{i\,\pp}{L^3}\, \Us{\Ls}, ~~ \im{\left[ \qf{\fv}, \qf{\fm} \right]}{\Lf} = i\,\Uf{\Lf}
\end{equation}
For the compact model, a Fourier transform with discrete Fourier modes for the Klein-Gordon system is at our disposal. Since the space adiabatic scheme requires the choice of an (arbitrary) discrete energy value of the Klein-Gordon system, it is self-evident to pass over to the mode description. Furthermore, we employ a particle description for the Klein-Gordon field i.e., expressing the Hamiltonian by means of creation- and annihilation operators for each Fourier mode. The Hamiltonian symbol is then given by, 
\begin{equation}
\pmb{h} = \EGsv\,\Uf{\Lf} + \frac{1}{\Sib} \sum_{\vec{k}\,\in\,\Theta} \ol \abd{k}{} \ab{k}{}
\end{equation}
where $\ol := \sqrt{k^2+\mu^2 b^2}$, and $\Theta := \frac{2 \pi}{\ell} \left(\mathbb{Z}^3\! \setminus\! \lbrace 0 \rbrace\right)$. 
The annihilation and creation operators satisfy the commutation relations,
\begin{equation}
\irm{\left[ \ab{k}{}, \abd{k}{'} \right]}{\FS} = \im{\delta}{\vec{k}, \vec{k}'}\,\Uf{\Lf},
\end{equation}
where $\im{\delta}{\vec{k}, \vec{k}'}$ is the Kronecker delta. The $\sv$-dependence of the creation- and annihilation operators will be examined in the next section. The representation of the commutation relations will be chosen as the tensor product, $\Hs{\Ls} \otimes \Hs{\Lf}$. The first factor is a simple $L^2$-space over the real axis with Lebesgue measure $\dr \sv$, while the second factor is the symmetric Fock space $\mathcal{F}_s(\ell^2(\Theta))$ with respect to the one particle Hilbert space $\ell^2(\Theta)$ of the discrete Fourier modes (counting measure understood).

\section{Space Adiabatic Perturbation Scheme}

\subsection{Parameter-Dependent Harmonic Oscillators}
We examine the characteristics of the Hamilton symbol, \eqref{eq:Hamilton Symbol}. The eigenvalue problem of the parameter-dependent Hamilton operator $\h{}{\scpb}$ for the Klein-Gordon subsector has the form,
\begin{equation}
\h{}{\scpb} \ef{\nd} = \Epd{n} \ef{\nd}.
\end{equation}
Here, $\nd$ is a short form for the number of excitations $\im{n}{\vec{k},d}$ for every wavenumber $\vec{k}$ and for the degeneracy label $d \in \lbrace 1,..., D \rbrace$, where $D$ is the multiplicity of the eigenenergy $\Epd{n}$. The $D$ degenerate eigensolutions $\ef{\nd}{\svb}$ are mutually orthonormal. The energy value is explicitely given by,
\begin{equation} \label{eq:Energy in Eigenband}
\Epd{n} =: \EGsv + \EKGsv{n} = \EGsv + \frac{1}{\Sib} \sum_{\vec{k}\in \Theta} \ol \im{n}{\vec{k},d}.
\end{equation}
The corresponding eigenstates are derived from the vacuum state $\Omega^{(\sv)} \in \mathcal{F}_s\left(\ell^2(\Theta)\right)$ as follows, 
\begin{equation}
\ef{\nd}= \prod_{\vec{k} \in \Theta} \frac{\left(\abd{k}{}\right)^{\nkd}}{\sqrt{\nkd!}} \Omega^{(\sv)}.
\end{equation}
For the procedure of  space adiabatic perturbation theory, we choose one particular eigenenergy $\Epd{n}$ and we denote the corresponding $\sv$-dependent projector as,
\begin{equation} \label{eq:Pi Zeroth Order}
\pib{n,0} := \sum_{d=1}^D \ef{\nd} \im{\left\langle \ef{\nd}, \cdot \right\rangle}{\Lf}
\end{equation} 
where $\im{\left\langle \cdot, \cdot \right\rangle}{\Lf}: \mathcal{F}_s\left(\ell^2(\Theta)\right) \times \mathcal{F}_s\left(\ell^2(\Theta)\right) \rightarrow \mathbb{C}$ denotes the inner product of the Klein-Gordon Hilbert space. Since we restrict the application to one particular, albeit arbitrary, eigenband with quantum number(s) $\nkd, \vec{k} \in \Theta$, we omit the index $n$ for the Moyal projector in what follows, i.e., we write $\pib{0} := \pib{n,0}$ instead. \\

The space adiabatic scheme uses the derivatives of the eigensolutions $\ef{\nd}$ with respect to the gravitational canonical pair. Since the eigensolutions do not depend on the momentum, $\sm$, it suffices to compute the $\sv$-derivative of $\ef{\nd}$. Similar to the simple quantum mechanical models in \cite{4}, it can be shown that, on the one hand, the $\sv$-derivative separately decreases the excitation number by two, for any wave number $\vec{m}$ which is already excited at least twice, i.e., for which $\en{m} \geq 2$. On the other hand, the $\sv$-derivative separately increases the excitation number by two for \emph{every} wave number $\vec{m}$. The twofold lowered and raised states enter with the respective factors,
\begin{equation}
\alpha_{1,\en{m}}^{\svb} := - \f \frac{\sqrt{(\en{m}\!-\!1)\,\en{m}}}{2}, ~~~ \alpha_{1,\en{m}}^{\svb} := - \f \frac{\sqrt{(\en{m}\!+\!1)(\en{m}\!+\!2)}}{2},
\end{equation}
where the function $\f$ is defined via the frequency,
\begin{equation}
\f := - \frac{1}{2} \frac{\partial \ln (\ol)}{\partial b}.
\end{equation}
Hereby, the $\sv$-derivative of $\ef{\nd}$ is given by,
\begin{align}
\frac{\partial \ef{\nd}}{\partial \sv} \!&=\! \sum_{\vec{m} \in \Theta} \prod_{\substack{\vec{k} \in \Theta \\ \vec{k}\neq \vec{m}}}\!\left( \alpha_{1,\en{m}}^{(b)} \frac{\left( \abd{k}{} \right)^{\en{k}}}{\sqrt{(\en{k})!}} \frac{\left( \abd{k}{} \right)^{\en{m}-2}}{\sqrt{(\en{m}-2)!}}+ \alpha_{2,\en{m}}^{(b)}  \frac{\left( \abd{k}{} \right)^{\en{k}}}{\sqrt{(\en{k})!}} \frac{\left( \abd{k}{} \right)^{\en{m}+2}}{\sqrt{(\en{m}+2)!}}\right)  \Omega^{(b)} \nonumber \\
&=: \sum_{\vec{m} \in \Theta} \left(\alpha_{1,\en{m}}^{(b)} \psi_{\lbrace..,\en{m}-2,..\rbrace}^{(b)} +\alpha_{2,\en{m}}^{(b)} \psi_{\lbrace..,\en{m}+2,..\rbrace}^{(b)} \right),
\end{align}
where $e_{\lbrace..,\en{m}\pm2,..\rbrace}^{\svb}$ denotes the state which is raised, respectively lowered, in the quantum number $\en{m}$ by two compared to $\ef{\nd}$.
\subsection{Structural Ingredients}
Space adiabatic perturbation theory requires three structural conditions from the model in order to be applicable.
\begin{itemize}
\item[1.] The quantum Hilbert space of the system decomposes as a tensor product, 
\begin{equation}
\mathcal{H}=\Hs{\Ls}\otimes \Hs{\Lf},
\end{equation}
and the dynamics in $\Hs{\Ls}$ happens on much larger scales as compared to the dynamics in $\Hs{\Lf}$. In this model, $\pp := \sqrt{\kappa/\lambda}$ represent the separation of these scales of change.  As argued in \cite{1}, this result is in line with the separation of the homogeneous degrees of freedom and the non-homogeneous field variables within the model.
\item[2.] Deformation quantization with the Weyl ordering is employable for the quantization of the homogeneous cosmological subsystem. This makes space adiabatic perturbation theory work on a technical level.
\item[3.]
The prinicipal symbol of the Hamiltonian $\h{}{\scpb}$, which is already the total Hamiltonian symbol for this model, has a pointwise isolated part of the spectrum $\im{\sigma}{0,\nd}^{\scpb}$. We choose one of the eigenspaces with energy label $\left\lbrace \nkd \right\rbrace$, $\vec{k} \in \Theta$. For fixed and distinct variables $\scpb$ an arbitrary shift in the quantum number(s) $\nkd$ produces a distinct energy value.
\end{itemize}
In the following, we work out the details of the space adiabatic scheme, which consists in,
\begin{itemize}
\item[1)] the construction of the Moyal projector, $\pipd{} \in S^{\infty}(\pp; \Ps{\Ls}, \mathcal{L}(\Hs{\Ls}))$
\item[2)] the construction of the Moyal unitary, $\us{}{\scpb} \in S^{\infty}(\pp; \Ps{\Ls}, \mathcal{L}(\Hs{\Ls}))$, and
\item[3)] the construction of the effective Hamiltonian, $\heff{}{\scpb} \in S^{\infty}(\pp; \Ps{\Ls}, \mathcal{L}(\Hs{\Ls}))$.
\end{itemize}

\subsection{Construction of the Moyal Projector $\pipd{n}$}
Space adiabatic perturbation theory rests on the space adiabatic theorem \cite{3}, which states that it is possible to construct iteratively a projection operator $\im{\qf{\Pi}}{(k)}$ of the full Hilbert space $\Hi$ up to order $k$ in the adiabatic perturbation parameter $\pp$, such that the subspace $\im{\qf{\Pi}}{(k)} \mathcal{H}$ is invariant under the evolution generated by the full Hamiltonian $\qs{\h{}{}}$. For further information on the scheme and first intuitive examples, we refer the reader to [REF,REF]. Here, we only state that space adiabatic perturbation theory suggests to construct the above projection operator on the symbol level, with the projection operator, \eqref{eq:Pi Zeroth Order} as the zeroth order starting point of the iterative construction scheme. The full Moyal projection operator has then the form of a formal perturbation series in $\pp$,
\begin{equation}
\pipd{n} := \sum_{\dl=1}^D \sum_{N=0}^{\infty} \pp^N\,\pipd{\nd,N}, ~~ \pipd{\nd,N} \in S^{\infty}(\Ps{\Ls}, \mathcal{L}(\mathcal{F}_s(\ell^2(\Theta))),
\end{equation}
and we recall that the index $\nd$ is a set of excitations number associated to the degeneracy label $\dl$. The index $n$ is then the shortcut for the set of all these excitations number for all degeneracy labels. \\
As explained more in detail in \cite{4,1}, the iterative conditions for the $N$-th order projection symbol, $\pipd{n,N}$, are given by,
\begin{enumerate}[label=\textnormal{\arabic*)}]
\item $\pipd{(N)}\,\im{\star}{\pp}\,\pipd{(N)}- \pipd{(N)}  = \mathcal{O}(\pp^{N+1})$, \label{itm:1}
\item $\left(\pipd{(N)}\right)^{\ast} - \pipd{(N)} = \mathcal{O}(\pp^{N+1})$, \label{itm:2}
\item $\im{[\h{}{}, \pipd{(N)} ]}{\im{\star}{\pp}} =  \mathcal{O}(\pp^{N+1})$, \label{itm:3}
\end{enumerate}
where ``$\im{\star}{\pp}$'' is the star product for the Weyl ordering, i.e., the pull back of the operator Weyl ordered multiplication on the space of semiclassical symbols. It is then straightforward to compute the first order contribution $\pipd{n,1}$ by means of the conditions \ref{itm:1}, \ref{itm:2}, \ref{itm:3} and the zeroth order projector, \eqref{eq:Pi Zeroth Order}. Thereby, we define the energy associated to a single excitation with respect to the mode $\vec{k}$ as, $\DE{k} := \ol/\Sib$. As a shorthand notation, we denote the eigenstate $e_{\lbrace..,\en{k}\pm2,..\rbrace}^{(b)}$, which is raised, respectively lowered in the quantum number $\en{k}$ by two compared to $\ef{\nd}$, by $\ef{\nd \pm 2}$. Then, the contribution of first order to the Moyal projector is given by,
\begin{align}
\pli{1} = \frac{i}{2\,L^3} \sum_{d=1}^D \sum_{\vec{k}\in \Theta}& \left( \aone{k} \Eone \left( \prora{-2} - \prola{-2}  \right) \right. \\
&~~~+\left. \atwo{k} \Etwo \left( \prora{+2} - \prola{+2}  \right) \right) 
\end{align}
where we defined the $n$-dependent functions,
\begin{align}
\Eone &:= \left( \frac{1}{\im{\Delta}{E,\vec{k}}} \left( \frac{\partial \EG}{\partial \sm} - \frac{1}{\Sib} \frac{\partial \Sib}{\partial \sm} \EKGr \right) + \frac{1}{\Sib} \frac{\partial \Sib}{\partial \sm}\right),\\
\Etwo &:= \left(- \frac{1}{\im{\Delta}{E,\vec{k}}} \left( \frac{\partial \EG}{\partial \sm} - \frac{1}{\Sib} \frac{\partial \Sib}{\partial \sm} \EKGr \right) + \frac{1}{\Sib} \frac{\partial \Sib}{\partial \sm}\right),
\end{align}
with $\EG$, $\EKGr$ and $\Sigma$ respectively defined in \eqref{eq:Hamilton Symbol}, \eqref{eq:Energy in Eigenband} and \eqref{eq:Definition slow variables}. We emphasize that the Weyl quantization of the projector $\pipd{(1)}$ projects on a subspace of the full Hilbert space which is $\pp$-dependent, and the description of the dynamics therein is non-trivial. Space adiabatic perturbation theory therefore suggests to construct a Moyal unitary symbol $\us{}{\scpb} \in S^{\infty}(\pp;\Ps{\Ls}, \mathcal{L}(\mathcal{F}_s(\ell^2(\Theta)))$ which maps the dynamics of $\pipd{}$ to a suitable reference space $\Hs{0}$. This is the aim of the next section.

\subsection{Construction of the Moyal Unitary $\us{n}{\scpb}$}
For the given model, the simplest and physically most convenient choice of a reference space for projecting the dynamics from $\im{\mathcal{H}}{\text{KG},n}^{\scpb} := \pipd{(1)}\Hs{\Lf}$ on, is given by taking $\pib{0} \Hs{\Lf}$ for one particular $\sv = \svz$. In this section, we denote it as $\Hs{0}$. The corresponding ``reference'' projector in $\Hs{\Lf}^{\scpb}$ will be denoted by,
\begin{equation}
\pli{\refs} := \sum_{d=1}^D \proR
\end{equation}

In order to mediate between $\Hs{\Lf}^{\svb}$ and $\Hs{0}$, and vice versa, a unitary operator $\us{}{}$ is necessary. The condition of unitarity and the requirement that $\us{0}{}$ should map $\pli{0}$ to $\pli{\refs}$ gives at least the following conditions on $\us{0}{}$,
\begin{enumerate}[label=\textnormal{\arabic*)}]
\item $\us{0}{}\,\cdot\,\pli{0}\,\cdot(\us{0}{})^{\ast} = \pli{\refs}$,
\item $\us{0}{}\,\cdot (\us{0}{})^{\ast} = \Uf{\Hs{0}}$,
\item $(\us{0}{})^{\ast} \cdot \us{0}{} = \Uf{\Lf}$.
\end{enumerate}
Therefore, we employ for $\us{0}{}$ the following operator-valued symbol,
\begin{equation}
\us{0}{\svb} = \sum_{j \geq 0} \im{e}{j}^{(\svz)} \im{\left\langle \ef{j}, \cdot \right\rangle}{\Lf},
\end{equation}
where the index $j$ is a short notation for the set of all possible excitation configurations within the Fock space $\mathcal{F}_s(\ell^2(\Theta)))$. This choice trivially satisfies the conditions on $\us{0}{}$ and is simple and evident.\\

Taking $\us{0}{}$ and the above conditions as a starting point, we aim to construct iteratively a semiclassical symbol $\us{\rel}{\scpb} \in S^{\infty}(\pp; \Ps{\Ls}, \mathcal{L}(\Hs{\Lf}))$. The formal power series has the form,
\begin{equation}
\us{\rel}{\scpb} = \sum_{N \geq 0} \pp^N \us{\rel,N}{\scpb}, ~~~ \us{\rel,N}{\scpb} \in S^{\infty}(\mathcal{L}(\Hs{\Lf})).
\end{equation}
Transcription of the conditions for $\us{0}{}$ using the $\im{\star}{\pp}$-product, gives for the semiclassical symbol $\us{}{} \in S^{\infty}(\pp; \Ps{\Ls},\mathcal{L}(\Hs{\Lf}))$,
\begin{enumerate}[label=\textnormal{\arabic*)}]
\item $\us{}{}\; \im{\star}{\pp}\; \pli{}\; \im{\star}{\pp}\; \left(\us{}{}\right)^{\ast} = \pli{\refs}$
\item $\us{}{}\;\im{\star}{\pp}\; (\us{}{})^{\ast} = \Uf{\Hs{0}}$,
\item $(\us{}{})^{\ast}\;\im{\star}{\pp}\; \us{}{} = \Uf{\Lf}$.
\end{enumerate}
Like for the Moyal projector, the perturbative equations for $\us{}{}$ read, when considered order by order in $\pp$,
\begin{enumerate}[label=\textnormal{\arabic*)}]
\item $\us{(N)}{}~ \im{\star}{\pp}~ \pli{}{}~ \im{\star}{\pp}~ \left(\us{(N)}{}\right)^{\ast}- \pli{\refs} = \mathcal{O}(\pp^{N+1})$,
\item $\us{(N)}{}~\im{\star}{\pp}~ \left(\us{(N)}{}\right)^{\ast} -\Uf{\Hs{0}} =  \mathcal{O}(\pp^{N+1})$,
\item $\left(\us{(N)}{}\right)^{\ast}~ \im{\star}{\pp}~ \us{(N)}{} - \Uf{\Lf} =  \mathcal{O}(\pp^{N+1})$.
\end{enumerate}
Since for the computation of the effective Hamilton symbol of second order, only the first order of the unitary symbol is necessary, we restrict our analysis to the computation of $\us{1}{\scpb}$. Given $\us{0}{}$, it is straightforward to show that the hermitian part of $\us{1}{}$ vanishes because $\us{0}{}$ is independent of $\sm$. The remaining anti-hermitian part is given by, 
\begin{align}
\im{\pmb{u}}{1} = \frac{i}{2\,L^3} \sum_{d=1}^D \sum_{\vec{k}\in \Theta}& \left( \aone{k} \Eone \left( \prorf{-2} + \prolf{-2}  \right) \right. \\
&~~~+\left. \atwo{k} \Etwo \left( \prorf{+2} + \prolf{+2}  \right) \right) 
\end{align}

\subsection{Construction of the Effective Hamiltonian $\heffscp{,n}$}

The last step of the perturbation scheme consists in pulling the dynamics of the chosen subspace to the $\pp$-independent subspace, $\Hs{0}= \im{\qs{\qf{\Pi}}}{\refs} \Hi$. This essentially means that by applying the unitary operator $\qs{\qf{u}}$ which is the Weyl quantization  on the Hamiltonian $\qs{\h{}{}}$, the action of the latter on elements in $\qs{\qf{\Pi}} \Hi$ is rotated to $\Hs{0}$.  We denote the semiclassical symbol,
\begin{equation} \label{eq:hsymbol_equation}
\heff{}{} := \us{}{} ~ \im{\star}{\pp}~ \h{}{} ~ \im{\star}{\pp}~ \left(\us{}{}\right)^{\ast},
\end{equation} 
as the effective Hamiltonian.
Then, the Weyl quantization, $\irm{\qs{\qf{h}}}{eff}$, of the symbol $\heff{}{} \in S^{\infty}(\pp;\Ps{\Ls}, \mathcal{L}(\Hs{\Lf}))$ is essentially self-adjoint on the Schwartz space $\mathcal{S}(\mathbb{R}, \Hs{\Lf})$. And in particular, the dynamics of $\qs{\h{}{}}$ are mapped unitarily to $\Hs{0}$, such that,
\begin{align}
\left[ \im{\qs{\qf{h}}}{\mathrm{eff}}, \im{\qs{\qf{\pi}}}{\refs} \right] &=0, \\
e^{-i \qs{\qf{H}}s} - \left(\qs{\qf{u}}\right)^{\ast} e^{-i\,\irm{\qs{\qf{h}}}{eff}\,s}\, \qs{\qf{u}} &= \mathcal{O}(\pp^{\infty} \left| s \right|), \label{eq:Effh_Almost_Invariance}
\end{align}
where $s\in\mathbb{R}$ is a real parameter.\\

We construct $\heff{}{}$ perturbatively by means of equation \eqref{eq:hsymbol_equation} up to second order. We assume for the generic form of the semiclassical symbol $\heff{}{}$,
\begin{equation}
\heff{}{\scpb} = \sum_{N \geq 0}^2 \pp^N\, \heff{,N}{\scpb},~~ \heff{,N}{\scpb} \in S^{\infty}(\Ps{\Ls}, \mathcal{L}(\Hs{\Lf}))
\end{equation}
Its restriction up to the $N$-th order, $\heff{,(N)}{}$ is defined as,
\begin{equation} \label{eq:OM_heffCondk}
\heff{,(N)}{} = \us{(N)}{}\,\im{\star}{\pp}\,\h{(N)}{}\,\im{\star}{\pp}\, \left(\us{(N)}{}\right)^{\ast} + \mathcal{O}(\pp^{N+1}).
\end{equation} 
Since we are mainly interested in the effective dynamics within the chosen subspace associated to $\im{\qf{\pi}}{\refs}$, we directly restrict the effective Hamilton symbol on this subspace by multiplying our results for $\heff{,(2)}{}$ by $\im{\qf{\pi}}{\refs}$ from the left and the right. We denote the latter symbol then by $\heff{,n,(2)}{}$.\\

At zeroth order of the perturbation theory, the effective Hamilton symbol is then given by,
\begin{align}
\heff{,n,0}{} &= \irm{\pmb{\pi}}{\refs}\cdot \im{\pmb{u}}{0}\cdot\pmb{h}\cdot\im{\pmb{u}}{0}^{\ast} \cdot \irm{\pmb{\pi}}{R} \\
&= \left(L^3 \left( - \frac{1}{12} \Sigma\,\frac{\sm^2}{\sv^2} + \Lambda\Sigma^3 \right) + \frac{1}{\Sigma} \sum_{\vec{k}\in \Theta} \ol \nkd \right) \irm{\pmb{\pi}}{R}.
\end{align}
This corresponds to the Born-Oppenheimer adiabatic limit of the perturbation theory in which the effective Hamiltonian for the gravitational degrees of freedom not only contains the first ``bare'' gravitational part $\EGsv$, but also the backreaction contribution from the Klein-Gordon energy band $\nd$ that has been chosen.\\

The first order effective Hamiltonian symbol $\heff{,n,1}{}$, which can be computed straightforwardly using \eqref{eq:OM_heffCondk} together with the result for $\heff{,n,0}{}$, has no contribution within the chosen subspace. However, the generic effective Hamiltonian symbol does not vanish and it enters in the computation of the next order effective Hamiltonian symbol contribution. The space adiabatic perturbation scheme yields \emph{a priori} for the second order contribution of the effective Hamiltonian, 
\begin{align}
\heff{,n,2}{} = &\sum_{d=1}^D \left( \sum_{\vec{m}\in \Theta} \left(\frac{\im{E}{\mathrm{eff},3}^{\scpb}}{(\im{\omega}{m}^{\scriptscriptstyle{(\sv)}})^3} \left( \im{n}{\vec{m},d} + \frac{1}{2}\right) + \frac{\im{E}{\mathrm{eff},4}^{\scpb}}{(\im{\omega}{m}^{\scriptscriptstyle{(\sv)}})^4} \left(\im{n}{\vec{m},d}^2+ \im{n}{\vec{m},d} + 1\right) + \frac{\im{E}{\mathrm{eff},5}^{\scpb}}{(\im{\omega}{m}^{\scriptscriptstyle{(\sv)}})^5} \left( \im{n}{\vec{m},d} + \frac{1}{2}\right) \right) \right) \nonumber \\
& \cdot \im{e}{\im{n}{d}}^{(\im{\sv}{0})} \langle \im{e}{\im{n}{d}}^{(\im{\sv}{0})}, \cdot \rangle \label{eq:EffHam2}
\end{align}
where we employed the energy functions,
\begin{align}
\im{E}{\mathrm{eff},3}^{\scpb} &:= \frac{1}{8\,\ell^6}\left( \ol \frac{\partial \,\ol}{\partial b} \right)^2 \left(\frac{1}{\Sib^3} \left( \frac{\partial \Sib}{\partial \sm}\right)^2 - \frac{1}{\Sib^2} \left( \frac{\partial^2 \Sib}{\partial \sm^2} \right)\right) = -\frac{\sigma^2 \mu^4 b^2}{8\, \Sib^5} \\
\im{E}{\mathrm{eff},4}^{\scpb} &:= \frac{1}{16\,\ell^6}\left( \ol \frac{\partial \,\ol}{\partial b} \right)^2 \left( \frac{2}{\Sib} \frac{\partial \Sib}{\partial \sm} \frac{\partial \EG}{\partial \sm} + \frac{\partial^2 \EG}{\partial \sm^2} - \frac{1}{\Sib} \frac{\partial^2 \Sib}{\partial \sm^2} \EKGr \right)\\
\im{E}{\mathrm{eff},5}^{\scpb} &:= \frac{1}{8\,\ell^6}\left( \ol \frac{\partial \,\ol}{\partial b} \right)^2\left(- \Sib \left(\frac{\partial \EG}{\partial \sm} \right)^2 + 2 \frac{\partial \Sib}{\partial \sm} \frac{\partial \EG}{\partial \sm} \EKGr - \frac{1}{\Sib} \left( \frac{\partial \Sib}{\partial \sm} \right)^2 \EKGr^2 \right).
\end{align}
Note that these functions do not depend on the wave vector $\vec{m}$ which has been employed as a summation index in \eqref{eq:EffHam2}. They act as multiplicative functions which could be pulled out of the sums. The explicit evaluation of the energy functions shows that several terms include higher orders in the perturbation parameter $\varepsilon$. The remaining terms at second order are,
\begin{align} \label{eq:heffective second order}
\heff{n,2}{} = - \frac{3\,\mu^4}{32} \sum_{d=1}^D &\im{e}{\im{n}{d}}^{(\im{\sv}{0})} \langle \im{e}{\im{n}{d}}^{(\im{\sv}{0})}, \cdot \rangle \\
&\cdot \sum_{\vec{m}\in\Theta} \left( \frac{b^4}{\ell^3 \Sib^3} \frac{1}{(\im{\omega}{m}^{\scriptscriptstyle{(\sv)}})^4} \left(\im{n}{\vec{m},d}^2+ \im{n}{\vec{m},d} + 1 \right) + \frac{3 \sm^2 \sv^2}{\Sib} \frac{1}{(\im{\omega}{m}^{\scriptscriptstyle{(\sv)}})^5} \left( \im{n}{\vec{m},d} + \frac{1}{2}\right) \right). \nonumber
\end{align}
We emphasize that the sums over all modes $\vec{m}$ in \eqref{eq:heffective second order} converge. First, the integers $\im{n}{\vec{m},d}$ are only non-vanishing for a finite number of modes $\vec{m}$ which solves the convergence problem for terms which enter with polynomials of $\im{n}{\vec{m},d}$. The remaining constant contributions however benefit from the high inverse order with which, $\im{\omega}{m}^{\scriptscriptstyle{(b)}} = \sqrt{{\vec{m}}^2 + \mu^2 b^2}$, enters.

\section{Conclusion and Outlook}
\label{s4}

We have computed an effective Hamiltonian that incorporates the influence 
of the inhomogeneous degrees of freedom on the quantum dynamics of the 
homogeneous ones. We have done this for every energy band of the 
Hamiltonian of the inhomogeneous degrees of freedom separately. 
The spectrum of these effective Hamiltonians can be computed and 
by rotating the corresponding (generalised) eigenvectors by the (approximate) 
inverse unitary operator that was used to achieve the (approximate) 
adiabatic decoupling of the energy bands, one obtains (approximate) 
eigenvectors of the original Hamiltonian that describes the 
interaction and mutual backreaction 
between the homogeneous and inhomogeneous degrees of freedom.
One can then consider semiclassical states and decompose them 
with respect to this (approximate) generalised energy basis in order 
to study their quantum evolution and in particular the fate of the 
classical big bang singularity. We reserve this for future work.

In the final paper \cite{15} 
of this series we consider General Relativity without dust 
coupled to an inflaton. We start from the formulation of the dynamics in 
terms of the canonical variables \cite{16} which are already ideally prepared 
for an application of the SAPT scheme. The challenge is twofold:
First, the dependence of the inhomogenous contribution to the Hamiltonian 
constraint on the homogeneous degrees of freedom is more complicated than 
for the model treated in this paper which 
makes the computation of the adiabatic corrections much more complicated. 
Second, the avoidance of the complications originating from the subset of 
the slow phase space where the Mukhanov-Sasaki and tensor mode mass squared 
functions become negative requires a more involved discussion. \\
\\
\\
{\bf Acknowledgements}\\
\\
S.S. thanks the
Heinrich-B\"oll Stiftung for financial and intellectual support and the
German National Merit Foundation for intellectual support.

}

\end{document}